\documentclass[prd,preprintnumbers,showpacs,onecolumn,showkeys,floatfix,nofootinbib,notitlepage]{revtex4-1}
\usepackage{graphicx}
\usepackage{amssymb}
\usepackage{amsmath}
\usepackage{mathtools}
\usepackage{epsfig}
\usepackage{color}
\usepackage{enumerate}
\usepackage{slashed}
\usepackage{hyperref}
\usepackage{epstopdf}

\def\slashchar#1{\setbox0=\hbox{$#1$}
   \dimen0=\wd0 \setbox1=\hbox{/} \dimen1=\wd1
   \ifdim\dimen0\big>\dimen1 \rlap{\hbox to \dimen0{\hfil/\hfil}} #1
   \else  \rlap{\hbox to \dimen1{\hfil$#1$\hfil}} / \fi}

\newcommand{\ud}{\mathrm{d}}
\newcommand{\be}{\begin{equation}}
\newcommand{\ee}{\end{equation}}
\newcommand{\bea}{\begin{eqnarray}}
\newcommand{\eea}{\end{eqnarray}}

\newcommand{\Appendix}[1]%
    {%
     \section{#1}%
      }

\begin{document}

\title{Equivalence Between the Gauge $n\cdot\partial n\cdot A=0$ and the Axial Gauge}

\author{Gao-Liang Zhou}
\email{ zhougl@itp.ac.cn}
\affiliation{College of Science, Xi'an University of Science and Technology, Xi'an 710054, People's Republic of China}
\author{Zheng-Xin Yan}
\affiliation{College of Science, Xi'an University of Science and Technology, Xi'an 710054, People's Republic of China}
\author{Xin Zhang}
\affiliation{College of Science, Xi'an University of Science and Technology, Xi'an 710054, People's Republic of China}




\begin{abstract}
Discontinuity of gauge theory in the gauge condition $n\cdot\partial n\cdot A=0$, which emerges  at $n\cdot k=0$, is studied here. Such discontinuity is different from that one confronts in axial gauge and can not be regularized by conventional analytical continuation method. The Faddeev-Popov determinate of the gauge $n\cdot\partial n\cdot A=0$, which is solved explicitly in the manuscript, behaves like a $\delta$-functional of gauge potentials once singularities in the functional integral is neglected and the length along $n^{\mu}$ direction of the space tends to infinity. As a sequence, perturbation series in the gauge $n\cdot\partial n\cdot A=0$ returns to that in axial gauge for short-range correlated objects that are free from singularities in path integral. However, the equivalence between the gauge $n\cdot\partial n\cdot A=0$ and axil gauge is nontrivial for  long-range correlated objects and quantities that are affected by singularities in path integral. Continuity of gauge links one encounter in perturbation theory and lattice calculation is affected by such discontinuity.
\end{abstract}


\maketitle

Although be of great importance in topics related to non-Abelian gauge theory,  the Faddeev-Popov quantization\cite{Faddeev:1967fc} of non-Abelian gauge theory is disturb by the famous Gribov ambiguity \cite{Gribov-1978,Singer-1978}. Such flaw is  ubiquitous in non-Ableian gauge theory on $3$-sphere($S^{3}$) and $4$-sphere($S^{4}$) given that the gauge group is compact.\cite{Singer-1978}, which is different from quantization of Abelian gauge theory like quantum electrodynamics(QED).  As a fundamental issue, studies on the Gribov ambiguity are crucial for understanding infrared aspects quantum chromodynamics(QCD) and may be helpful to conquer the confinement  problem.

Gribov copies related to infinitesimal gauge transformations can be eliminated through the method of Gribov region, in which the Faddeev-Popov operator is positive definite\cite{Gribov-1978,Vandersickel:2012tz}. The Gribov region was constructed through the no pole condition\cite{Gribov-1978,Sobreiro:2004us} initially, which requires that non trivial poles of ghost propagator should be exclude from the Gribov region.   The Gribov region can also be constructed trough the famous Gribov-Zwanziger(GZ) action\cite{Zwanziger:1988jt,Zwanziger:1989mf,Zwanziger:1992qr}. These two methods are equivalent to each other\cite{Capri:2012wx}. Propagator of gluons in the Gribov region vanishes in infrared region in Landau gauge, which is different from traditional propagators of massless particles.  To be consistent with lattice data, however, propagator of gluons in Landau gauge should be nonzero in infrared region\cite{Cucchieri:2007md,Bogolubsky:2007ud}. It seems that refinement of gluon propagator in Gribov region is necessary once dynamical effects of auxiliary fields in  GZ action are taken into account\cite{Dudal:2007cw,Dudal:2008sp}. Authors in \cite{Su:2014rma} apply the CZ gluon propagator to the hot quark-gluon plasma and  get a new massless excitation of of quarks. In \cite{Guimaraes:2015vra}, a quark confinement model inspired by GZ action is presented, which possesses  nontrivial thermodynamic properties at finite temperature.

The Gribov ambiguity, however, does not vanish even if one works in Gribov region. Instead of Gribov region, it seems more reasonable to work in the fundamental modular region(FMR)\cite{Vandersickel:2012tz,vanBaal:1991zw}, in which the functional
\begin{equation}
\int\ud^{4}x tr[A^{U}_{\mu}(x)A^{U}_{\mu}(x)]
\end{equation}
takes absolute minima value,  where $U$ represents arbitrary gauge transformations.
However, it is difficult to realize  such procedure analytically.
Authors in \cite{Serreau:2012cg,Serreau:2013ila} present a method to average over Gribov copies with suitable weights, which is free from
the famous Neuberger zero problem in ordinary Fadeev-Popov quantization. It seems that analytical calculations in the scheme are rather complicated.

In \cite{zhou:2016gb}, we present a new gauge condition for non-Abelian gauge theory in the direct product space of straight line and $3-$torus($R\bigotimes (S_{1})^{3}$), which reads
\begin{equation}
n\cdot \partial n\cdot A(x)=0,
\end{equation}
where $n^{\mu}$ is the directional vector along $x_{i}-$axis($i=1,2,3$). It is proved that the gauge condition is free from Gribov ambiguity except for configurations of which the integral measure is zero. In addition, the gauge condition is continuous for more general configurations compared with the axil gauge. The Lehmann representation and canonical commutation relation are subtle if $n^{\mu}$ is time like\cite{West:1982gg}. We do not consider the case here.

The space $R\bigotimes (S_{1})^{3}$ is topologically equivalent to that one confront in quantum mechanics in the box normalization scheme.  Topological properties of the space may be related to the confinement phenomenon\cite{tHooft:1981sps}. We notice that the famous BCS superconductivity is caused by the electron-phonon interaction at low temperature\cite{Bardeen:1957mv}. It is not surprising that contributions of  quarks and gauge potentials with non-trivial crystal wave vectors may be related to confinement(``superinsulator"). While considering the Gribov ambuguity, which one does not confront in QED, we do not consider gauge potentials with non-trivial topology  for simplicity. Gauge potentials with non-trivial crystal wave vectors and gauge transformations with torsion will be studied in other works.

Generally speaking, one can not choose a continuous gauge transformations $U$ in the space $R\bigotimes (S_{1})^{3}$ so that $n\cdot A^{U}=0$ as eigenvalues of the Wilson line
\begin{equation}
\label{Wilsonline}
W\equiv P\exp(ig\int_{-\frac{L}{2}}^{\frac{L}{2}}\ud s n\cdot A(x^{\mu}+sn^{\mu}))
\end{equation}
are invariant under continuous gauge transformations in the space $R\bigotimes (S_{1})^{3}$, where $n^{\mu}$ is the directional vector along $x_{i}-$axis($i=1,2,3$) and $L$ represents the scale of the space along $n^{\mu}$ direction.  However, as displayed in \cite{zhou:2016gb}, one can choose a continuous gauge transformation $U$ so that $n\cdot\partial n\cdot A^{U}=0$ given that generator of the Wilson line (\ref{Wilsonline}) is continuous in the space $R^{4}$. It seems that contributions of the mode
\begin{equation}
\label{mode}
\int_{-\frac{L}{2}}^{\frac{L}{2}}\ud n\cdot x n\cdot A(x)
\end{equation}
can not be removed by continuous gauge transformations in the space $R\bigotimes (S_{1})^{3}$. Contribution of such mode cause the discontinuity of tree level gluon propagator in the gauge  $n\cdot\partial n\cdot A=0$ as displayed in \cite{zhou:2016gb}.

In this paper, we consider contributions of the mode (\ref{mode}) and show the discontinuity of the theory at $n\cdot k=0$. One confronts such singularity in axial gauge as gluon propagator in axial gauge is singular for $n\cdot k=0$. However, such singularity is caused by gluons with infinitesimal $n\cdot k$ and does not affect physical quantities. For the theory considered here, the discontinuity is caused by contributions of gluon modes with $n\cdot k=0$. Tree level gluon propagator is discontinuous at  $n\cdot k=0$ as displayed in \cite{zhou:2016gb}. Such result is extended to higher orders here.

Without loss of generality, we choose $n^{\mu}$ as
\begin{equation}
n^{\mu}=(0,0,0,1)
\end{equation}
in following calculations.
The Faddeev-Popov determinate of the gauge condition $n\cdot \partial n\cdot A=0$ reads\cite{zhou:2016gb},
\begin{equation}
\label{functionaldet}
Det(n\cdot\partial(n\cdot\partial-ign\cdot A))
=(\prod_{l=-\infty}^{\infty}\frac{2\pi l}{L})(\prod_{m> 0,\lambda}\frac{4\pi^{2} m^{2}}{L^{2}}-g^{2}\lambda^{2}),
\end{equation}
where $\lambda$ stands for  eigenvalue of $n\cdot A(x)$. Null eigenvalues of $n\cdot\partial$ have been dropped as they can be eliminated by gauge transformations independent of $n\cdot x$.  We see that the operator $(n\cdot\partial-ign\cdot A)$ is singular for
 \begin{equation}
 \label{degeneracy}
 \lambda=\pm\frac{2\pi m}{gL}\quad (m>0).
 \end{equation}
In fact $n\cdot A(x)$ is an antisymmetric hermitian matrix in adjoint representation. As a result, $-\lambda$ is eigenvalue of $n\cdot A(x)$ given that $\lambda$ is eigenvalue of $n\cdot A(x)$. If $\lambda$ satisfy (\ref{degeneracy}), then we have
 \begin{equation}
 e^{ig\lambda L}=e^{-ig\lambda L},\quad \lambda\ne -\lambda.
 \end{equation}
Gauge potential configurations that satisfy above equation  form the region in which the gauge condition  $n\cdot \partial n\cdot A=0$ suffers from the Gribov ambiguity\cite{zhou:2016gb}. This confirms the result in \cite{zhou:2016gb} that the gauge condition  $n\cdot \partial n\cdot A=0$ is free from the Gribov ambiguity except for  configurations with zero integral measure.

Determinate of the operator $-i n\cdot \partial$ is independent of fields and can be dropped. Relevant part of (\ref{functionaldet})  reads,
\begin{eqnarray}
\label{relevantdet}
Det(in\cdot\partial+gn\cdot A)
&=&(\prod_{m> 0,\lambda}\frac{4\pi^{2} m^{2}}{L^{2}}-g^{2}\lambda^{2})
\nonumber\\
&=&\mathcal{C}\left (\prod_{\lambda}\frac{2\sin(\frac{g\lambda L}{2})}{gL\lambda}\right ),
\end{eqnarray}
where $\mathcal{C}$ represents a constant independent of gauge potentials and can be dropped. To obtain the result, we have made use of the formula,
\begin{equation}
\frac{\sin x}{x}=\prod_{n=1}^{\infty}(1-\frac{x^{2}}{n^{2}\pi^{2}}).
\end{equation}
It is interesting to consider the asymptotic behavior of the determinate for $L\to\infty$. At first sight, one has
\begin{equation}
\label{asympotic}
\frac{2\sin(\frac{g\lambda L}{2})}{gL\lambda}\rightarrow\frac{\pi}{L}\delta(\frac{g\lambda}{2})\quad \text{(for $L\to\infty$)}.
\end{equation}
Therefore one may conclude that the gauge condition $n\cdot \partial n\cdot A=0$ is equivalent to axial gauge for $L\to\infty$. To examine the equivalence, we consider an arbitrary smooth function $f(x)$ and the integral
\begin{equation}
\label{intrgral}
\int_{-\infty}^{\infty}\ud x f(x)\frac{2\sin(\frac{gx L}{2})}{gLx}=\frac{1}{gL}\int_{-\infty}^{\infty}\ud x f(x)\int_{-\frac{gL}{2}}^{\frac{gL}{2}}\ud k e^{ikx}.
\end{equation}
If the integral is well defined then we can change the order of the integral and have,
\begin{eqnarray}
\int_{-\infty}^{\infty}\ud x f(x)\frac{2\sin(\frac{gx L}{2})}{gLx}&=&\frac{1}{gL}\int_{-\frac{gL}{2}}^{\frac{gL}{2}}\ud k f(k)
\nonumber\\
&\to& \frac{2\pi}{gL}f(0)\quad \text{(for $L\to\infty$)}.
\end{eqnarray}
Thus the relation (\ref{asympotic}) is valid once the integral (\ref{intrgral}) is well defined. Similarly, the gauge condition $n\cdot \partial n\cdot A=0$ is equivalent to axial gauge for $L\to\infty$ once the functional integral is well defined.

However, practical functional integrals are disturbed by various singularities, such as ultraviolet divergences and mass singularities. Let us consider the problem in the frame of perturbation theory.  In perturbation theory, integral over gauge potentials are controlled by exponent of action. Roughly speaking, the gauge potential mode with momentum $k^{\mu}$ is of order $(k^{2})^{1/2}$. Thus the integral is well defined given that $k^{2}\ne 0$. However, for the case $k^{2}=0$, which is related to mass singularity, the integral may be divergent. Thus the equivalence between the gauge condition $n\cdot \partial n\cdot A=0$ and axial gauge is violated unless mass singularities do not affect the quantity one concerned.

To specify above discussion, we consider the vacuum expectation value of an arbitrary operator $\mathcal{O}(x)$, which reads,
\begin{equation}
T\big<\mathcal{O}(x)\big>=
\frac{\int[\mathcal{D}A]\mathcal{O}(x)
e^{iS}\delta(n\cdot\partial n\cdot A)
\left (\prod_{\lambda}\frac{2\sin(\frac{g\lambda L}{2})}{gL\lambda}\right )}
{\int[\mathcal{D}A]
e^{iS}\delta(n\cdot\partial n\cdot A)
\left (\prod_{\lambda}\frac{2\sin(\frac{g\lambda L}{2})}{gL\lambda}\right )}.
\end{equation}
While concerning perturbation theory, free part of the action is kept in the exponent and interaction part is expanded.
We are interested in asymptotic behaviour of contributions of the modes with $k^{2}=0$. Therefore, we consider contributions of the special region
\begin{eqnarray}
n\cdot A(x)-n\cdot A_{k}(x)&\sim& \epsilon n\cdot A_{k}(x)
\nonumber\\
n\cdot A_{k}(x)&\equiv&\frac{1}{V}\int\ud^{4} y n\cdot A(y)e^{ik\cdot y} e^{-ik\cdot x}
\nonumber\\
k^{\mu}&=&(k,k,0,0),
\end{eqnarray}
where $\epsilon$ is an infinitesimal and $V$ is the volume of the space. This does not affect the result if the equivalence between the gauge condition $n\cdot \partial n\cdot A=0$ and axial gauge does hold for this case.     We consider the case that the operator can be written as polynomial of $n\cdot A_{k}(x)$ and write then integrals over $n\cdot A_{k}(x)$ as,
\begin{equation}
\int [\mathcal{D}n\cdot A_{k}] f(n\cdot A_{k})\left (\prod_{\lambda}\frac{2\sin(\frac{g\lambda L}{2})}{gL\lambda}\right ),
\end{equation}
where $f(n\cdot A_{k})$ is a polynomial of $n\cdot A_{k}(x)$. If the order of $f(n\cdot A_{k})$ is high enough, then the integral is divergent. This is  in contradict with the equivalence between the gauge condition $n\cdot \partial n\cdot A=0$ and axial gauge. We conclude that the equivalence can be violated by contributions of modes related to mass singularities.

As is well known, perturabtaive expansion is a kind of asymptotic expansion.  Thus the functional integral may be singular even if all  terms of perturbaitve series are well defined. Especially contributions of classical field configurations may cause singularities in functional integral. We consider the special case that some classical field configurations form a continuous function space with infinite volume. According to the principal of least action, the action of  classical field configuration should be smaller than that of configurations near the classical field configuration. Therefore the action is a constant in the continuous space of classical field configurations. As a result, integral over such space is not controlled by the exponent. That is to say, the equivalence between the gauge condition $n\cdot \partial n\cdot A=0$ and axial gauge may break down at non-perturbative level.

We notice that the gauge condition $n\cdot \partial n\cdot A=0$ is equivalent to axial gauge for gauge potential modes with $n\cdot k\ne 0$. Thus the break down of  equivalence signals the discontinuity of the theory at $n\cdot k=0$.  In \cite{zhou:2016gb}, such discontinuity is discussed at tree level though explicit calculation of tree level gluon propagator. Above discussions display that the discontinuity is related to singularities in path integral at all perturbative orders and may has non-perturbative effects.   For example, we consider the Wilson line
\begin{equation}
W(x)\equiv P\exp(ig\int_{0}^{x^{3}}\ud s n\cdot A(x^{0},x^{1},x^{2},s).
\end{equation}
Generally speaking, one has
\begin{equation}
W(x^{\mu}+Ln^{\mu})\neq W(x)
\end{equation}
according to the non-equivalence between the gauge condition $n\cdot \partial n\cdot A=0$ and axial gauge. Such discontinuity may affect lattice calculations, in which gauge links are necessary to maintain the obvious gauge invariance of the theory(see, for example, Ref.\cite{Kogut:1982ds}).

While working in the frame of perturbation theory, one often deal with quantities be free from mass singularities, like cross sections which are inclusive enough. One may wonder that wether such quantities are  affected by the discontinuity of the theory at $n\cdot k\ne 0$. To clarify the problem, we modify the Lagrangian density according to the manner
\begin{equation}
\mathcal{L}(x)\to \mathcal{L}(x)+\frac{i}{2}m^{2}A^{\mu}A^{\mu}(x),
\end{equation}
where $m$ is a small quantity. The modified Lagrangian density is not gauge invariant or Lorentz invariant and may destroy the unitarity.  However,  the defect is not harmful in the limit $m\to 0$ if the functional integral is well defined for $m=0$. After the modification, the exponent in path integral becomes
\begin{equation}
e^{iS}\to e^{iS}e^{-\frac{1}{2}m^{2}\int\ud^{4}x A^{\mu}A^{\mu}(x)}.
\end{equation}
Integral over gauge potentials is controlled by the term $e^{-\frac{1}{2}m^{2}\int\ud^{4}x A^{\mu}A^{\mu}(x)}$ for $m\ne 0$. Thus the the gauge condition  $n\cdot \partial n\cdot A=0$ is equivalent to axial gauge for $m\ne 0$ and $L\to \infty$. If the path integral is well defined for $m\to 0$ then the equivalence is valid even for $m\to\infty$. Thus the equivalence valid is in the limit $L\to \infty$ if and only if the quantity one concerned is free from singularities in functional integral.

To see the accuracy of the equivalence for quantities free from singularities in path integral, we consider the propagator of gauge particles,
\begin{eqnarray}
G(k)&=&\int\ud^{4} x e^{ik\cdot x}T\big<n\cdot A(x)n\cdot A(0)\big>.
\end{eqnarray}
To simplify the calculation, we consider $n\cdot A(x)$ as number not matrix in color space. This is enough for the estimation of the propagator here. We work in Euclidian space according to Wick rotation and have,
 \begin{eqnarray}
G(k)&=&
\int\ud^{4}x e^{ik\cdot x}
\frac{\int[\mathcal{D}A]n\cdot A(x)n\cdot A(0)
e^{-S}\delta(n\cdot\partial n\cdot A)
\left (\prod_{\lambda}\frac{2\sin(\frac{g\lambda L}{2})}{gL\lambda}\right )}
{\int[\mathcal{D}A]
e^{-S}\delta(n\cdot\partial n\cdot A)
\left (\prod_{\lambda}\frac{2\sin(\frac{g\lambda L}{2})}{gL\lambda}\right )}
\nonumber\\
&=&
\int\ud^{4} xe^{ik\cdot x}
\frac{\int_{-gL}^{gl} \int[\mathcal{D}\phi]\int[\mathcal{D}A]n\cdot A(x)n\cdot A(0)
e^{-S}e^{\frac{i}{a^{3}L}\int\ud^{4}x n\cdot A(x)\phi(x)}\delta(n\cdot\partial n\cdot A)}
{\int_{-gL}^{gl} \int[\mathcal{D}\phi]\int[\mathcal{D}A]
e^{-S}e^{\frac{i}{a^{3}L}\int\ud^{4}x n\cdot A(x)\phi(x)}\delta(n\cdot\partial n\cdot A)}
\nonumber\\
&\sim & \frac{-i}{k^{2}}\frac{L^{3}}{a^{3}}e^{-\frac{g^{2}L^{2}}{a^{2}}},
\end{eqnarray}
where $a$ represents the integral measure of each point in space time and can be related to lattice distance in lattice calculations.
In above calculation, we have assumed that lengths along all directions of the space are of the same order. We see that the equivalence between the gauge condition $n\cdot \partial n\cdot A=0$ and axial gauge is of high high accuracy for $k^{2}\ne 0$. Thus perturbative calculations based on axial gauge and the gauge presented here are equivalent to each other for short distance quantities free from singularities in path integral.

In conclusion, we have calculated the Faddeev-Popov determinate of the gauge condition $n\cdot\partial n\cdot A=0$ explicitly.  After the summation over ghost loops, the gauge condition presented here behaves like the axial gauge. It is proved through explicit calculations that such equivalence is valid only for quantities free from singularities in path integral. For short distance quantities which are free from mass singularises, the equivalence works well in the level of perturbation theory. Non-equivalence between the two gauge conditions may cause discontinuity of the theory in the gauge $n\cdot\partial n\cdot A=0$ at $n\cdot k=0$. For Abelian gauge theory, the covariance gauge is continuous and free from the Gribov ambiguity. Therefore it is reasonable to believe that the discontinuity does not affect gauge invariant objects in Abelian gauge theory. For non-Abelian gauge theory, however, the situation is subtle and more researches are necessary. Especially definition of asymptotic quark and gluon states may be affected by such discontinuity.

\section*{Acknowledgments}
We thank for Professor Y. Q. Chen and Doctor Z. Y. Zheng for helpful discussions and important suggestions on the manuscript. The work of G. L. Zhou is supported by The National Nature Science Foundation of China under Grant No. 11647022 and The Scientific Research Foundation for the Doctoral  Program of Xi'an University of Science and Technology under Grant No. 6310116055 and The Scientific Fostering Foundation of Xi'an University of Science and Technology under Grant No. 201709. The work of Z. X. Yan is supported by  The Department of Shanxi Province Natural Science Foundation of China under Grant No.2015JM1027.

\bibliography{discontinuity}

\end{document}